\documentclass{PoS}

\usepackage{epsfig}
\usepackage{latexsym}
\usepackage{graphicx}
\usepackage{bm}

\renewcommand{\vec}[1]{\boldsymbol{#1}}

\title{A novel computation of the thermodynamics of the SU($3$) Yang-Mills theory}

\ShortTitle{Thermodynamics of the SU($3$) Yang-Mills theory}

\author{\speaker{Leonardo Giusti}\\
        Dipartimento di Fisica, Universit\`a di Milano-Bicocca\\
        and INFN, sezione di Milano-Bicocca\\
        Edificio U2, Piazza della Scienza 3\\ 
        20126 Milano, Italy.\\
       E-mail: \email{Leonardo.Giusti@mib.infn.it}}

\author{Michele Pepe\\
        INFN, Sezione di Milano-Bicocca\\ 
        Edificio U2, Piazza della Scienza 3\\ 
        20126 Milano, Italy.\\
        E-mail: \email{Michele.Pepe@mib.infn.it}}

\abstract{We present an accurate computation of the Equation of State of the SU(3)
Yang-Mills theory using shifted boundary conditions in the temporal direction. In this
framework, the entropy density $s$ can be obtained in a simple way from the expectation value of the
space-time components $T_{0k}$ of the energy-momentum tensor. At each given value of the
temperature, $s$ is measured in an independent way at several values
of the lattice spacing. The extrapolation to the continuum limit shows small discretization effects
with respect to the statistical errors of approximatively 0.5$\%$.}

\FullConference{The 33rd International Symposium on Lattice Field Theory\\
                 14 -18 July  2015\\
                 Kobe International Conference Center, Kobe, Japan}

\begin{document}
\section{Introduction}
The pressure, $p$, the energy density, $e$, and the entropy density $s$, are main
features of Quantum Chromo Dynamics (QCD) at finite temperature $T$. The Equation of State describes
the temperature dependence of the above quantities, and it is of crucial relevance in many
areas. It is an important input in the analysis of data collected at the heavy-ion
colliders, in the study of nuclear matter, and in astrophysics and cosmology when strongly
interacting matter is under extreme conditions. Many collaborations have put a lot of
effort in calculating the thermodynamics features of QCD by numerical simulations on the
lattice, both in the pure gauge sector and with dynamical fermions. These studies are
challenging from the numerical viewpoint: $p$, $e$, and $s$ are related to the
free energy density which suffers from an ultraviolet additive power-divergent renormalization,
and cannot be directly measured in a Monte Carlo simulation.

In Ref.~\cite{Boyd:1996bx} a first accurate investigation of the Equation of State of
the SU($3$) Yang-Mills theory was accomplished by numerical simulations on the lattice. 
This approach has turned out to be very successful; however it has the drawback that a
subtraction at $T=0$, or at some other temperature \cite{Borsanyi:2012ve}, has to be
performed. Interestingly, there are also other equations that relate the pressure, the
energy density and the entropy density to the expectation values of the matrix elements
of the energy-momentum tensor $T_{\mu\nu}$ ~\cite{Landau1987}. A practical use of those
equations in Monte Carlo simulations on the lattice, however, requires the computation of the
renormalization constants of the bare lattice
tensor~\cite{Caracciolo:1989pt,Giusti:2011kt,Giusti:2012yj}.  

The energy-momentum tensor contains the currents associated to Poincar\'e symmetry and
scale transformations. As a consequence, when the regularization of a quantum theory
preserves space-time symmetries --~like, for instance, dimensional regularization~--
$T_{\mu\nu}$ does not renormalize.
The lattice regularization explicitly breaks the Poincar\'e invariance, which is recovered
only in the continuum limit. Thus, the bare energy-momentum tensor 
needs to be properly renormalized to guarantee that the associated charges generate
translations and rotations in the continuum limit~\cite{Caracciolo:1989pt}.
For the Yang-Mills theory, scale invariance is also broken by the regularization; however, that
symmetry is anomalous and it is not restored in the continuum limit, generating a dynamical mass-gap.

The proper approach to define non-perturbatively the renormalized energy-momentum tensor on
the lattice is to impose the validity of some Ward Identities at fixed lattice spacing up
to terms that vanish in the continuum limit~\cite{Caracciolo:1989pt}. Based on that framework, the
renormalization constants of $T_{\mu\nu}$ have been computed in perturbation theory at
1 loop~\cite{Caracciolo:1991cp}. Although one can in principle construct a lattice
definition of the energy-momentum tensor, the non-perturbative calculation of the
renormalization factors can be not straightforward if one has to consider correlation
functions that are difficult to measure by numerical simulations. 

A few years ago, a thermal quantum field theory has been formulated in a moving reference
frame using the path-integral language~\cite{Giusti:2010bb,Giusti:2011kt,Giusti:2012yj}. This
setup can be implemented by considering a spatial shift ${\vec \xi}$ when closing the
boundary conditions along the temporal direction. The shift ${\vec \xi}$ corresponds to the
Wick rotation of the speed of the moving frame. Interestingly, in this new framework, one
can write down new Ward Identities involving the energy-momentum tensor, that allow to
compute in a simple way the non-perturbative renormalization factors of the energy-momentum tensor
on the lattice~\cite{Giusti:2012yj,Giusti:2015daa}. Furthermore, since the shift breaks explicitly the
parity symmetry, there are new equations relating the thermodynamic quantities
and the expectation values of off-diagonal matrix element of $T_{\mu\nu}$.
Numerical simulations with shifted boundary conditions have already provided new successful,
simple methods to study the thermodynamics of the Yang-Mills theory~\cite{Giusti:2014ila}. 

This report is organized as follows. In section 2, the main equations with shifted
boundary conditions are summarized both in the continuum and on the lattice. The next
section presents the results of the non-perturbative calculation of the renormalization
factors of the energy-momentum tensor and, in section 4, the renormalized
energy-momentum tensor is used to compute the Equation of State of the SU($3$) Yang-Mills
theory. Conclusions and outlook follow.

\section{Ward Identities with shifted boundary conditions}
We consider the thermal SU($3$) Yang-Mills theory in the Euclidean space in the path-integral formulation
with shifted boundary conditions~\cite{Giusti:2012yj}
\begin{equation}\label{eq:shfbc}
A_\mu(L_0,\vec x) =A_\mu(0,\vec x - L_0\vec\xi) \;,
\end{equation}
where $L_0$ is the system size along the compact direction and $\vec \xi \in\mathbb{R}^3$
is the shift. When $\vec \xi \neq 0$, the parity symmetry is broken, and there are new
interesting Ward Identities involving the energy-momentum tensor
$T_{\mu\nu}$~\cite{Giusti:2010bb,Giusti:2011kt,Giusti:2012yj} ($x_0\neq 0$) 
\begin{equation}\label{eq:dxi}
L_0 \langle \; T_{0k} \rangle_{\vec\xi} = \frac{1}{V}
\frac{\partial}{\partial \xi_k} \ln Z(L_0,\vec \xi)\;, \qquad
L_0 \langle \; {\overline T}_{0k}(x_0) \, O(0) \rangle_{\vec\xi,\, c}
= \frac{\partial}{\partial \xi_k} \langle O \rangle_{\vec\xi}\;.
\end{equation}
The functional $Z(L_0,\vec \xi)$ is the partition function with shifted boundary conditions, and $O$
is a generic gauge invariant operator. The subscript $c$ 
indicates a connected correlation function, $\langle \cdot \rangle _{{\vec\xi}}$ stands
for the expectation value with shifted boundary conditions, and ${\overline T}_{\mu\nu}=\int d^3x\,
T_{\mu\nu}(x)$. The field $T_{\mu\nu}$ can be defined by
\begin{equation}
T_{\mu\nu} (x) = \frac{1}{g_0^2} 
\left[ F_{\mu\rho}^a (x) F_{\nu\rho}^a (x)  
-\frac{1}{4} \delta_{\mu\nu} F_{\rho\sigma}^a (x) F_{\rho\sigma}^a (x) \right]\; , 
\end{equation}
where $g_0$ is the bare coupling constant. The field strength is given in terms of the
gauge field $A_\mu (x)$ by 
$F_{\mu\nu}(x) = \partial_\mu A_\nu (x) - \partial_\nu A_\mu (x) -i [A_\mu (x),A_\nu (x)]$.
Other useful equations are the following~\cite{Giusti:2012yj,Giusti:2015daa}
\begin{equation}\label{eq:WIodd}
\langle T_{0k} \rangle_{\vec\xi} = \frac{\xi_k}{1-\xi_k^2} 
\left\{\langle T_{00} \rangle_{\vec\xi}  
- \langle T_{kk} \rangle_{\vec \xi}\,\right\}\; ,
\qquad
\frac{\partial}{\partial \xi_k} \langle T_{\mu\mu} \rangle_{\vec\xi} =
\frac{1}{(1+\xi^2)^2}\frac{\partial}{\partial \xi_k}
\left[\frac{(1+\xi^2)^3}{\xi_k} \langle T_{0k} \rangle_{\vec\xi} \right]\; . 
\end{equation}
It is important to note that the above equations involve the expectation value of the
off-diagonal matrix element $ \langle T_{0k} \rangle$, which may be non vanishing due
to the breaking of parity symmetry.

When we consider the lattice regularization, the 10-dimensional symmetric $SO(4)$ representation
of the energy-momentum tensor splits into the sum of the sextet, the triplet, and the singlet 
irreducible representations of the hyper-cubic group. The field
$T_{\mu\nu}$ can then be expressed as a combination of the following three operators
\begin{equation}\label{eq:lattmunu}
T^{[1]}_{\mu\nu} = (1-\delta_{\mu\nu}) \frac{1}{g_0^2} F^a_{\mu\alpha}F^a_{\nu\alpha}\; , \;\;
T^{[2]}_{\mu\nu} = \delta_{\mu\nu}\, \frac{1}{4 g_0^2}\, F^a_{\alpha\beta} F^a_{\alpha\beta}\; ,\;\;
T^{[3]}_{\mu\nu} = \delta_{\mu\nu} \frac{1}{g_0^2} \Big\{F^a_{\mu\alpha}F^a_{\mu\alpha} 
- \frac{1}{4} F^a_{\alpha\beta}F^a_{\alpha\beta} \Big\}\; , 
\end{equation}
and of the identity. Since translation and rotation symmetries are broken by the lattice,
the sextet $T^{[1]}_{\mu\nu}$ and the triplet $T^{[3]}_{\mu\nu}$ operators
pick up a multiplicative renormalization factor only, while the singlet $T^{[2]}_{\mu\nu}$
mixes also with the identity.
The renormalized energy-momentum tensor can finally be written as
$ T^{\rm R}_{\mu\nu} = Z_{_T} \Big\{T^{[1]}_{\mu\nu} + z_{_T}  T^{[3]}_{\mu\nu} + 
z_{_S} \big[T^{[2]}_{\mu\nu} - 
\langle T^{[2]}_{\mu\nu} \rangle_0 \big]\Big\}$ .

Because of the finite renormalization factors, we can write the lattice version of the first
equation of (\ref{eq:dxi}) as follows
\begin{equation}\label{ZTpractic}
Z_{_T}(g_0^2) = - \frac{\Delta f}{\Delta \xi_k}\, \frac{1}{\langle T^{[1]}_{0k} \rangle_{\vec\xi} }\; , 
\qquad \mbox{with} \quad
\frac{\Delta f}{\Delta \xi_k} = \frac{1}{2a V}
\ln\Big[\frac{Z(L_0,{\vec\xi }-a \hat k/L_0 )}{Z(L_0,{\vec\xi}+a \hat k/L_0 )} \Big] .
\end{equation}
The two equations in (\ref{eq:WIodd}) become
\begin{equation}\label{Zd}
z_{_T}(g_0^2) = \frac{1-\xi_k^2}{\xi_k} \frac {\langle T^{[1]}_{0k}
  \rangle_{\vec\xi} }{\langle T^{[3]}_{00}\rangle_{\vec\xi} - \langle T^{[3]}_{kk} \rangle_{\vec\xi}}\; ,
\end{equation}
\begin{equation}\displaystyle\label{eq:singlet}
z_{_S} = \frac{1}{(1+\xi^2)^2}\frac{\left[\frac{(1+\xi^{'2})^3}{\xi'_k} \langle T^{[1]}_{0k} \rangle_{\vec\xi'} 
\right]_{\vec\xi'=\vec\xi+ a \hat k/L_0} - 
\left[\frac{(1+\xi^{'2})^3}{\xi'_k} \langle T^{[1]}_{0k} \rangle_{\vec\xi'} 
\right]_{\vec\xi'=\vec\xi- a \hat k/L_0}
}
{\langle T^{[2]}_{\mu\mu} \rangle_{\vec\xi+ a \hat k/L_0} - \langle T^{[2]}_{\mu\mu} \rangle_{\vec\xi- a \hat k/L_0}}\; . 
\end{equation}
Note that the equations (\ref{ZTpractic})--(\ref{eq:singlet}) allow for a fully non-perturbative
definition of $T_{\mu\nu}$. They also suggest simple procedures to perform the numerical
calculations of $Z_{_T}$, $z_{_T}$ and $z_{_S}$ since only the expectation values of local
fields need to be measured. 

\section{The non-perturbative renormalization factors}
In this section we present the results of Monte Carlo simulations to calculate $Z_{_T}
(g_0^2)$ and $z_{_T} (g_0^2)$ in the range $g_0^2\in(0,1)$~\cite{Giusti:2015daa}; work is in progress
for the computation of  $z_{_S} (g_0^2)$. We define the SU($3$) Yang--Mills theory on a
space-time lattice of volume $L_0\times L^3$ and lattice spacing $a$. We impose periodic boundary conditions in the
spatial directions and shifted boundary conditions in the compact one:
$U_\mu(L_0,\vec x) =U_\mu(0,\vec x - L_0\vec\xi)$, where $U_\mu(x_0,\vec x)$ are the link variables. 
We consider the standard Wilson action $S[U] = -1/g_0^2\, \sum_{x,\mu\nu} {\rm Re}\mbox{Tr} [ U_{\mu\nu}(x) ]$,
where the plaquette is given by
$U_{\mu\nu}(x) = U_\mu(x)\, U_\nu(x+ a\hat \mu)\, U^\dagger_\mu(x + a\hat \nu)\,
U^\dagger_\nu(x)$. The gluon field strength tensor is defined as~\cite{Caracciolo:1989pt}  
\begin{equation}
F^a_{\mu\nu}(x) = - \frac{i}{4 a^2} 
\mbox{Tr} \Big\{\Big[Q_{\mu\nu}(x) - Q_{\nu\mu}(x)\Big]T^a\Big\}\; , 
\end{equation}
where $Q_{\mu\nu}(x) = U_{\mu\nu}(x) + U_{\nu-\mu}(x) + U_{-\mu-\nu}(x) + U_{-\nu\mu}(x)$,
and the minus sign stands for the negative orientation. The renormalization constants $Z_{_T}$, $z_{_T}$ and $z_{_S}$ are finite, 
and depend on $g^2_0$ only up to discretization effects.
Considering the above definition of the field strength tensor
on the lattice, at 1 loop in perturbation theory their expressions are~\cite{Caracciolo:1989pt,Caracciolo:1991cp}
\begin{equation}\label{eq:PT1loop}
Z_{_T}(g_0^2) = 1 + 0.27076 \; g_0^2\;, \quad
z_{_T}(g_0^2) = 1 - 0.03008\; g_0^2 \;, \quad 
z_{_S}(g_0^2) = \frac{b_0}{2} g_0^2\; . 
\end{equation}

\subsection{Computation of $Z_{_T}$\label{sec:Ztmeth}}
The direct determination of $\Delta f/\Delta \xi_k$ in Eq.~(\ref{ZTpractic}) is a numerically
challenging problem since it requires the computation of the ratio of two partition
functions with a poor overlap of the relevant phase space~\cite{deForcrand:2000fi,DellaMorte:2007zz,Giusti:2010bb}.  
\begin{figure}[thb]
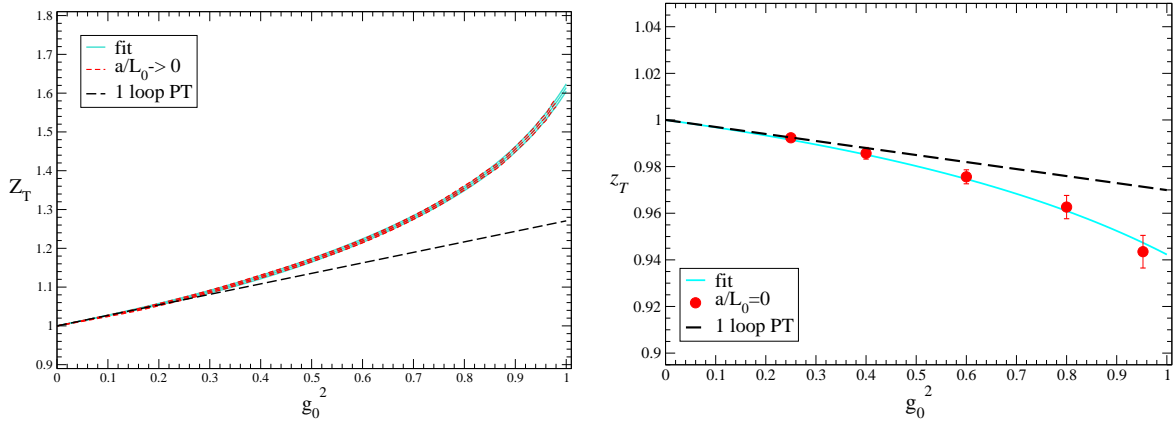

\hspace{-0.5cm}\includegraphics[width=7.5 cm,angle=0]{ZTfit.eps}\ \ \ \ \ 
\includegraphics[width=7.5 cm,angle=0]{zts.eps}
\caption{
The renormalization factor $Z_{_T}(g_0^2)$ (left panel) and $z_{_T}(g_0^2)$ (right panel)
as a function of the bare coupling $g_0^2$. The dashed lines represent the 1-loop
perturbative results, while the solid ones are interpolating fits of the numerical data.\label{fig:ZT}}
\end{figure}
Moreover, the calculation becomes quickly demanding for large lattices because the numerical
cost increases quadratically with the spatial volume. Since $\Delta f/\Delta \xi_k$ is a
smooth function of $g_0^2$ at fixed values of $L_0/a$ and $L/a$ in the range of chosen
values, its derivative with respect to $g_0^2$ can be written as  
\begin{equation}\label{intder}
\frac{d}{d g_0^{2}} \frac{\Delta f}{\Delta \xi_k} 
= \frac{1}{2a L^3 g_0^2}\,
\Big\{\langle S \rangle _{{\vec\xi} - a/L_0 \hat k} -\langle S \rangle _{{\vec\xi} + a/L_0 \hat k}\Big\}\; .
\end{equation}
The difference in the r.h.s.
has been computed for ${\vec\xi}=(1,0,0)$ and $L/a=48$ at $L_0/a=3,4$ and $5$ for many
values of $g_0^2$. At each value of $L_0/a$ the points are 
interpolated with a cubic spline, and the resulting curve is integrated over $g_0^2$. The 
free-case value is computed analytically and is added to the integral. Then $\langle
T^{[1]}_{0k} \rangle_{\vec\xi}$ has also been computed at many
values of $g_0^2$ and the results have been interpolated with cubic splines. The final
result for $Z_{_T}$ is shown in the left panel of Fig.~\ref{fig:ZT} together with the
1-loop perturbative result, and an interpolating fit gives
\begin{equation}\label{eq:ZTfinal}
Z_{_T}(g_0^2) = \frac{1 - 0.4457\, g_0^2}{1 - 0.7165\, g_0^2} - 0.2543\, g_0^4 
               + 0.4357\, g_0^6 - 0.5221\, g_0^8\; .
\end{equation}

\subsection{Determination of $z_{_T}$}
The renormalization constant $z_{_T}$ is calculated by 
imposing the tree-level improved version of Eq.~(\ref{Zd}) given by
$\{ z_{_T}(g_0^2) - \mbox{ free case}\}$ , with
$\frac{L\, \xi_k}{L_0(1+\xi_k^2)} = q \in \mathbb{Z}$. The expectation values of  
$\langle T^{[1]}_{0k} \rangle_{\vec\xi}$ and of the difference 
$\langle T^{[3]}_{00} \rangle_{\vec\xi} - \langle T^{[3]}_{kk} \rangle_{\vec\xi}$ are 
measured straightforwardly in the same simulation.

We chose ${\vec\xi} = (1/2,0,0)$ and $q=8$ so that 
the ratio of the spatial linear size over the temporal 
one is fixed to be $L/L_0=20$. We simulated 5 values of $g_0^2$ in the range 
$0\leq g_0^2 \leq 1$ with temporal length $L_0/a=4,6,8$ and $12$. After performing a
combined extrapolation to $a/L_0=0$, the final results are shown in the right panel of
Fig.~\ref{fig:ZT}. The dashed line is the 1-loop perturbative result, and the solid
one is an interpolating fit which gives
\begin{equation}\label{eq:ztsfinal}
z_{_T}(g_0^2) = \frac{1 - 0.5090\, g_0^2}{1 - 0.4789\, g_0^2}\; . 
\end{equation}

\section{The Equation of State}
In this section we use the non-perturbative calculation of the renormalization factors of the
energy-momentum tensor to obtain the Equation of State from Monte Carlo simulations. With
shifted boundary conditions, the entropy density $s$ can be written as~\cite{Giusti:2012yj}
\begin{equation}\label{EoS}
\frac{s}{T^3}=  -
\frac{L_0^4 (1+{\bf \xi} ^2)^3}{ \xi_k}  \langle T_{0k} \rangle_\xi Z_T,
\end{equation}
where $T=1/L_0\sqrt{1+{\bf \xi} ^2}$ is the temperature. In Ref.~\cite{Giusti:2014ila} the
temperature dependence of $s/T^3$ has been measured using the step-scaling
function: in that approach one can avoid computing $Z_T$ at all temperature values
, but only fixed constant steps in $T$ can be done. However, once the renormalization factor
$Z_T$ is known, the Eq.~(\ref{EoS}) allows to measure $s/T^3$ directly at any temperature. 

\begin{figure}[t!]
\hspace{-0.5cm}\includegraphics[width=7.5 cm,angle=0]{s_1.5Tc.eps}\ \ \ \ \ 
\includegraphics[width=7.875 cm,angle=0]{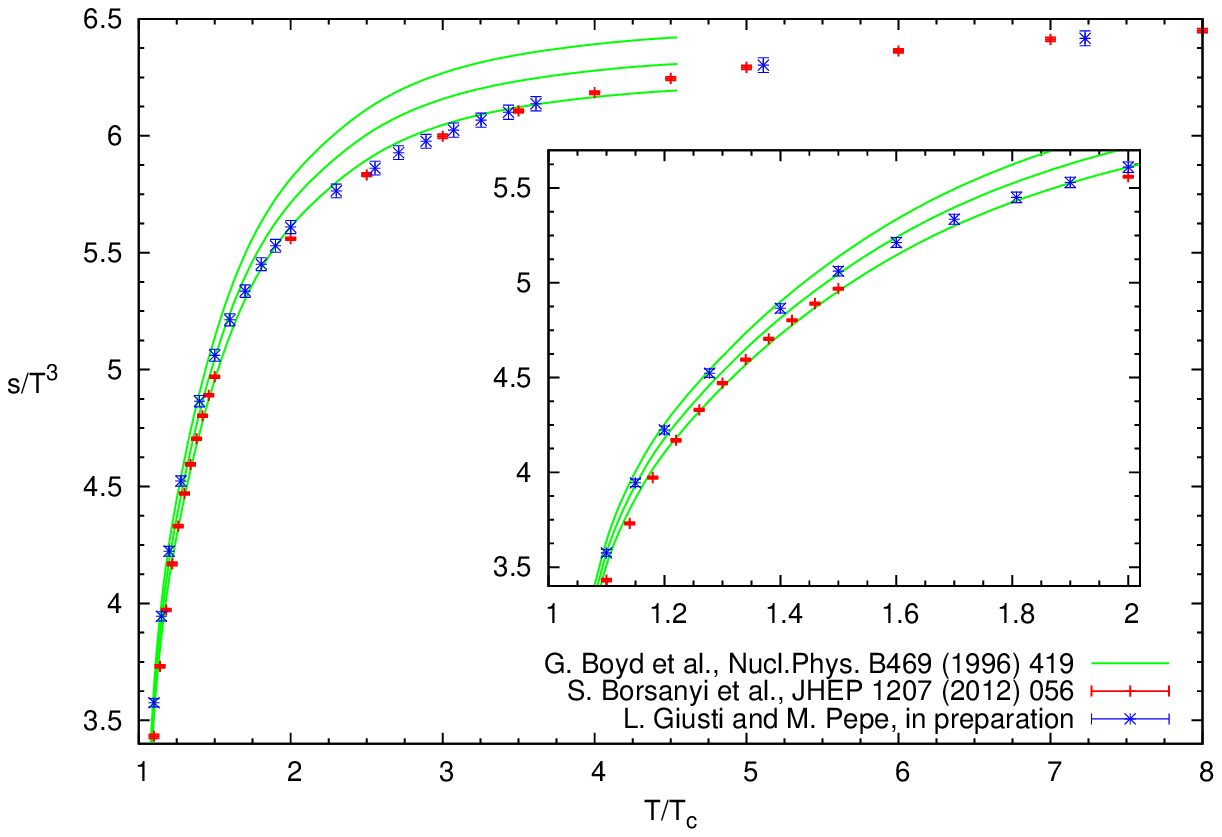}
\caption{Left: continuum limit extrapolation for $s/T^3$ computed as in Eq.~(\protect\ref{EoS}) at
$T=1.5 T_c$.
  Right: the temperature dependence of the dimensionless ratio $s/T^3$ in the continuum limit;
the results obtained using Eq.~(\protect\ref{EoS}) are compared with data available in the literature.\label{EoSplot}}
\end{figure}

We have performed numerical simulations at 21 different temperatures in the range between
$T_c$ and 7.5 $T_c$, where $T_c$ is the critical temperature of the 
theory. For each temperature we extrapolate $s/T^3$ to the continuum limit by considering
$L_0/a=5, 6, 7, 8$ and, sometimes, also $L_0/a=3, 4$ and 10. The extrapolation is performed
independently for every temperature, and we barely see
discretization effects within the numerical accuracy (see left plot of Fig.~\ref{EoSplot}).
The spatial volume has been chosen
to be $L/a=128$ for $L_0/a=3, 4, 5, 6$ and $L/a=256$ for $L_0/a=7, 8, 10$. Except for one point
with shift $\vec \xi =(1,1,1)$, we have always considered $\vec \xi =(1,0,0)$.
The scale is set using the Sommer scale~\cite{Sommer:1993ce,Necco:2001xg}, $r_0$, for $T<1.9\; T_c$ and
$L_{max}$ for higher temperatures~\cite{Capitani:1998mq}.

In Fig.~\ref{EoSplot} we compare our preliminary data with the results available in the
literature. In the region $1$-$2.5~T_c$ our data are compatible with those
in Ref.~\cite{Boyd:1996bx} that have significantly larger errors, while we find
a statistically significant discrepancy with the more precise ones in Ref.~\cite{Borsanyi:2012ve}.
At larger temperatures, up to $T\simeq 7\; T_c$, our results agree
with those presented in Ref.~\cite{Borsanyi:2012ve}. Work is in progress to clarify the
above mentioned discrepancy, and to reach temperatures of about $250~T_c$.

\section{Conclusions and outlook}
We have presented preliminary results of a new computation of the Equation of State
of the SU($3$) Yang-Mills theory. The computational strategy uses shifted
boundary conditions in the compact direction, and it can be applied in a straightforward
way to a generic SU($N$) Yang-Mills theory as well as to theories with dynamical
fermions. The approach relies on the non-perturbative calculation of the renormalization factors of the
energy-momentum tensor. The framework of shifted boundary conditions turns out to be very
effective and simple to investigate the Yang-Mills theory at finite temperature. Work is in progress
to include also dynamical fermions.

\end{document}